# Air-coupled ultrasound using broadband shock waves from piezoelectric spark igniters

K. G. Scheuer,[1] and R. G. DeCorby[2,a]

[1]*Ultracoustics Technologies Ltd., 40 Swallow Ave, Sherwood Park, AB, Canada, T8A 3H5*

[2] *ECE Department, University of Alberta, 9211-116 St. NW, Edmonton, AB, Canada, T6G 1H9*

We used an optomechanical sensor to study the ultrasound generated by manually operated piezoelectric spark igniters. These low-energy sparks produce short-duration acoustic shock-wave pulses, with sub-microsecond rise times and frequency content extending well beyond 2 MHz in air. The same source-receiver combination was then used to demonstrate broadband characterization of solid (polymer and glass) plates in a simple setup, where single spark events yielded high-SNR data without the need for critical alignment. This setup also enabled us to estimate pressure excursions approaching $10^5$ Pa at millimeter-scale distances from the spark. The results are in large part made possible by the small size, wide bandwidth, and high sensitivity of the optomechanical sensor, and might be of interest for air-coupled ultrasound applications in non-destructive testing.

---

[a] Author to whom correspondence should be addressed. Electronic mail: rdecorby@ualberta.ca

There is a long-standing interest in air-coupled ultrasound (ACUS)[1] for non-destructive testing (NDT), since it eliminates the need for water immersion or gel-coupling and enables rapid-throughput inspection of a diversity of industrial materials[2–10]. ACUS setups must typically contend with extreme insertion losses (>> 80 dB in some cases[1]), arising from air-solid impedance mismatches and the steep rise in air attenuation at MHz frequencies[11].

The most established approach to ACUS uses narrowband (*i.e.,* resonant) piezoelectric transducers, impedance-matched to air at a specific frequency[1,2,8]. Often the resonant frequencies of the transducer and sample must also be matched[1], which limits the flexibility of the setup. Broadband capacitive transducers have been used to study a variety of solid samples[4–7]. However, both require sophisticated (*i.e.* high-voltage and -frequency) drive and receive electronics, and their relatively large size makes them directional at MHz frequencies, such that alignment is critical[5]. Optical microphones[3] are a promising alternative when combined with a broadband source of ultrasound, for example generated using pulsed laser[3,9] or thermoacoustic[3,10] techniques.

Of related interest is the ultrasound produced by a low-energy spark discharge[12–20]. In its 'near field'[15], a spark produces a 'blast wave' pulse[17,20], which begins with a shock and a short compression phase followed by a longer rarefaction phase. At larger distances from the spark, the pulse can be approximated as an idealized 'N-wave'[13], with both an initial ('front') and final ('tail') shock. The front shock is characterized by a large (~ kPa) pressure increase on a time scale as short as ~ 0.1 $\mu$sec[14–17]. The pulse duration scales inversely with the discharge energy per unit length[12]; thus, lower-energy sparks produce higher-bandwidth acoustic signals, often extending into the MHz range. Experimental verification of such extreme pressure transients has been challenging, especially near the spark where the pressure excursions can greatly exceed the dynamic range of available instruments[13]. Moreover, capacitive microphones, and even optical interferometry methods[14–16], are typically unable to fully resolve the front-shock rise time. Here, we report that low-cost piezoelectric spark igniters[21] produce high-amplitude shock pulses (>$10^4$ Pa near the spark) with wider bandwidth content than for previously reported spark sources[13–19]. Moreover, we show that an air-coupled optomechanical microphone has sufficient bandwidth to resolve these pulses and sufficient sensitivity to detect them in transmission through a variety of plastic and glass plates.

We used a recently reported[22] optomechanical microphone, which provides high sensitivity in air (noise-equivalent pressure, NEP ~ 100 µPa/Hz$^{1/2}$) over a large bandwidth (several MHz). We have previously applied these devices to analysis and detection of low-pressure gas leaks[23] and to monitor the intrinsic thermal vibrations of small objects such as droplets[24] and bubbles[25]. Details on the particular sensor used here, including its calibration, are provided in the supplementary information (SI) file. We applied this sensor to the study of the acoustic pulses produced by several low-cost (~ $10 USD) piezoelectric spark igniters. These units employ a button-actuated mechanical impact to generate ~ 20 kV across a piezoelectric cell, producing low-discharge-energy (< 1 mJ)[21] sparks across mm-scale air gaps. Some of the results below used a particular brand of simple camp-stove igniters ('Optimus Sparky'), which generates a spark between a central pin and an outer barrel (~ 3 mm spark gap, see Fig. 1). For through-plate measurements, we used a barbecue replacement part ('Master Chef OEM Piezo Igniter') designed to generate a spark between externally wired electrodes and which we label as 'type B'. This device produced slightly higher sound pressure levels (~ 1.5x, for similar gap lengths) but with more experimentally observed strike-to-strike variability, both possibly attributable to a noticeably stiffer spring mechanism. Additional details on the devices and the experimental setups are provided in the SI file.

Figure 1(a) shows a representative time-domain pressure signal received at ~ 12 cm distance from a single spark generated by a type A device. This is as close as we could place our microphone to the spark while keeping both the positive and negative pressure excursions within its linear dynamic range. As is typical[13–16], the initial feature (highlighted by the dashed box) is a compression-rarefaction cycle intermediate between a blast-wave and an N-wave pulse[13]. The oscillations following this are attributable to acoustic reflections and diffraction within the microphone assembly[13,15,20]. The arrivals of reflected N pulses at delays of ~10, 20, and 40 µs corresponds to path delays of ~ 3-15 mm in air, close to the lateral dimensions of the microphone housing. Moreover, while the initial N-pulse was quite invariant with adjustments in the microphone angle (confirming the essentially omnidirectional characteristics of our sensors[22]), the remainder of the waveform was not. The trace shown was captured with the microphone at ~ 45 degrees to the spark direction. These reverberations can be reduced by appropriate baffling[13,14].

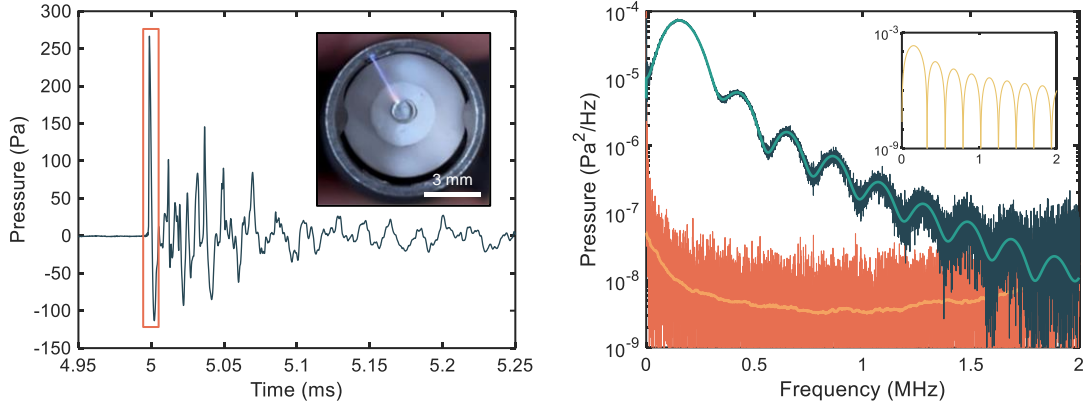

FIG. 1. (a) Typical time trace (at 12 cm spacing) of the pressure signal produced by a single spark event. The primary N-wave portion of the trace is indicated by the dashed box. The inset shows a typical spark produced by a type A device. (b) Non-averaged (single shot) energy spectrum of the primary N-wave pulse from part a. (blue) alongside the noise-floor (dark orange). The overlaid curves show the zero-padded N-pulse spectrum (green) and the averaged noise floor (light orange). The inset shows the theoretical spectrum for an ideal N-pulse with same peak pressure and half-duration of 2.2 µs (yellow).

The half-wave duration ($T < 2.5$ µs) of these pulses is substantially shorter (by a factor of at least 3-4x) than those reported previously for spark-generated shock waves[13–20], consistent with an inverse relationship between spark discharge energy and peak frequency of the generated ultrasound[12]. The piezo sparkers used have sub-millijoule discharge energies[21], more than an order of magnitude lower than that for typical capacitor-based discharge setups[13]. Figure 1(b) shows the corresponding frequency-domain spectrum for the *N*-wave portion (indicated by the dashed box in part (a)). Also shown is the energy spectrum predicted for an idealized *N*-wave given by[13]:

$$|P(f)|^2 = 4P_s^2 T^2 \{j_1(2\pi fT)\}^2 \quad , \quad (1)$$

where $P_s \sim 250$ Pa is the peak pressure excursion, $j_1$ is the spherical Bessel function of the first kind and first order, and $T = 2.2$ µs was used in the fit. The peak frequency at ~ 150 kHz is considerably higher than previously reported for spark-generated ultrasound[13–20], consistent with the short pulse length.

Given that the attenuation of ultrasound in air at 2 MHz is ~ 6.4 dB/cm[11], it is remarkable that content in this range is observed beyond 10 cm. This is partly attributable to the nonlinear propagation of the transient acoustic shock[13], which sharpens the pulse features and combats the low-pass filtering effect of air attenuation. Nevertheless, higher bandwidth and sharper features are expected nearer the spark. As shown in the SI file, the microphone could be biased to capture the front shock and initial compression phase at a distance as small as ~ 6 cm from the spark. This yielded a rise-time estimate of ~ 0.16 µs, in good agreement with theoretically predicted values for similar sparks[14,17]. Content extending beyond 5 MHz is present in the corresponding frequency domain plot.

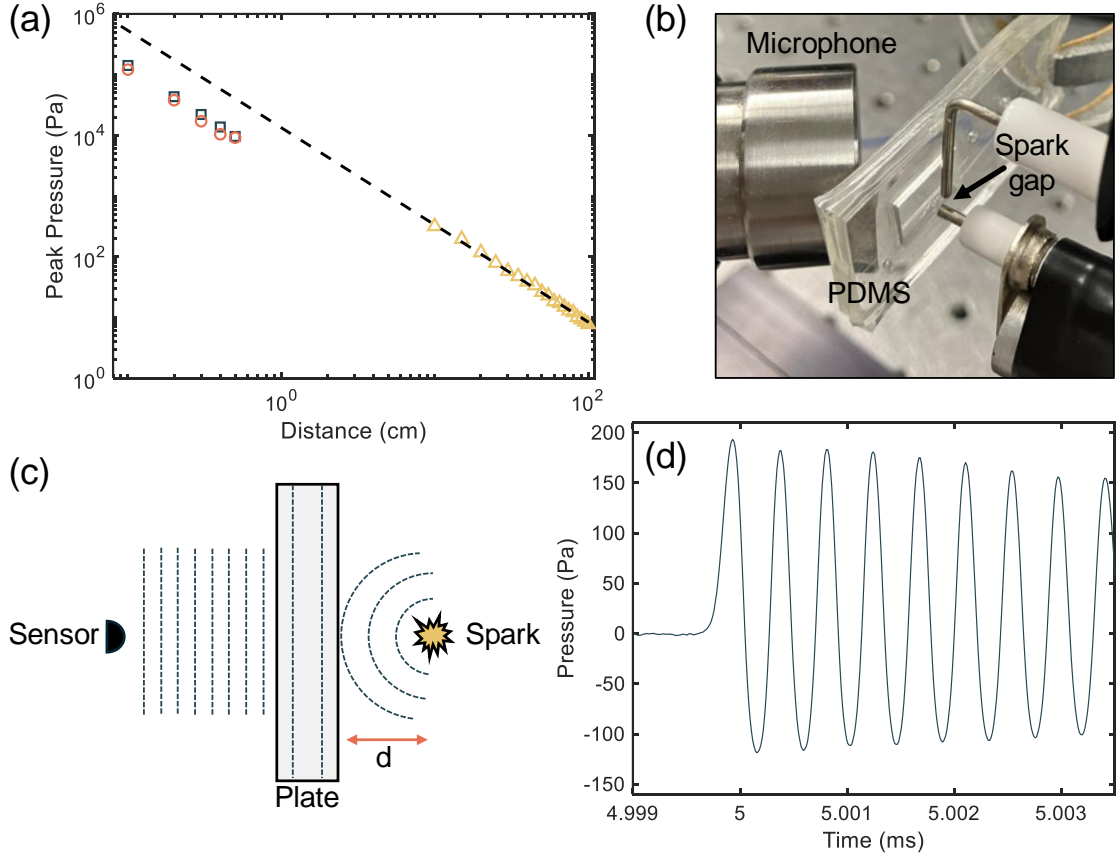

FIG. 2. (a) Peak positive pressure received at the microphone versus distance to the spark. At larger distances, the pressure was measured without an intervening PDMS layer (yellow triangles). The dashed line is a power-law fit to this data ($P_{max} \sim r^{-1.6}$). At shorter distances, the pressure was measured with either ~ 4 mm (orange circles) or ~ 2 mm (blue squares) of PDMS in the path, and the estimated peak pressure at the spark-side plate surface was plotted versus spacing between the spark gap and the PDMS (see main text). Each data point is an average of 6 measurements. (b) Photograph showing a PDMS sample (~ 2 mm thick 'window') between the spark gap and microphone. (c) Schematic illustration of the coupling of the spark blast-wave into a normal incidence compressional wave of an adjacent plate, a portion of which is transmitted to the small 'point' sensor aligned opposite the spark. (d) A typical time trace of the initial pressure signal received on transmission through the ~ 2 mm PDMS plate, showing a rise time on the order of 100 ns.

The spherical blast wave radiated by these low-energy sparks is expected to be within the limits of the so-called 'weak shock theory'[13–15,17]. It predicts that the peak pressure will exhibit power-law decay with distance; $P_{max}(r) \sim P_{max}(r_0) \cdot \exp(-\alpha \cdot r/r_0)$, where $r_0$ is a reference radial distance and $\alpha$ ~ 1.2 - 1.4 has typically been reported[13,17,20]. The yellow symbols in Fig. 2(a) indicate the peak positive excursion measured for distances in the 10-100 cm range. Each symbol is the average of multiple trials for a particular type A device, although the spark-to-spark variation was less than 5%, and similar pressure curves were measured for other devices (including type B devices with spark gap set to the same value of ~ 3 mm). The dashed curve is a power-law fit with $\alpha = 1.6$, representing slightly more rapid attenuation than for previously reported spark sources. We attribute this to the shorter pulse length and higher frequency content of these pulses, which increases the impact of air attenuation[13,14].

As in the present case, dynamic range issues have typically limited electrical-microphone-based studies of spark pressure fields to distances greater than ~ 10[13,14,17,18]. However, Karzova *et al.*[16] used an optical schlieren method to estimate peak pressures as high as ~ 12 kPa at 3 cm, and Liu *et al.*[15] used piezoelectric sensors to estimate a peak pressure as high as ~ 60 kPa at 2.5 cm from a higher-energy spark. Here, we used a thin sheet of PDMS as an 'acoustic attenuator' between the spark gap and the microphone, as shown in Fig. 2(b,c). Furthermore, we conservatively estimated the pressure produced by the spark as follows. First, we assume that the spherical blast wave acts as a purely normal force on the front surface of the plate and excites only a surface-normal longitudinal wave[19]. In reality, the divergent pressure field produced by the spark will also couple energy into off-normal shear and compressional waves[1,8,26]. Second, we assume that the optomechanical sensor predominately receives energy from the normal-incidence longitudinal wave. This is reasonable, since it is essentially a point sensor (~ 100 mm in size) placed opposite the spark at a distance > 5 mm from the plate. Moreover, even slightly off-normal wave components from the spark will tend to be totally reflected at the front plate surface[1], and any energy that is coupled into off-normal plate modes will tend to 'walk off' and be less efficiently radiated from the plate.

Based on these assumptions, we estimated the peak pressure at the front plate surface as the peak received pressure scaled by the normal-incidence reflection losses of the two air-PDMS interfaces. PDMS has an acoustic impedance of ~ 1 MRayl (see the SI file for further details), which implies a loss of ~ 28 dB at each air-PDMS interface. Thus, the first arrival of the blast-wave positive pressure excursion was assumed to be attenuated by 56 dB, corresponding to the pressure being reduced by a factor of ~ 650. We carried out experiments with two different thicknesses of PDMS and varied the spark-PDMS spacing in the 1 – 5 mm range. A typical result is shown in Fig.

2(d), with a peak received pressure of nearly 200 Pa in spite of the intervening PDMS plate. These signals displayed a rapid initial increase in pressure, with a rise time ~ 0.1 ms, in good agreement with theory[14]. The oscillations following the initial pulse have a period of ~ 0.4 ms and are due to excitation of the sensor fundamental resonance at ~ 2.5 MHz. This indicates that significant energy is delivered through the plate, extending to several MHz (see Fig. 4 below). For a range of spacings, the peak received pressure was multiplied by 650 and plotted in Fig. 2(a) versus spark-PDMS distance. The results were relatively insensitive to PDMS thickness, supporting the assumptions from above. This scaling is expected to underestimate the actual pressure, since it ignores all other sources of loss, including higher reflection at off-normal incidence, coupling to shear waves in the plate, as well as scattering and absorption.

It is interesting that these data points appear to follow the same power-law relationship, but with the estimated peak pressure lying a factor of 4-5x lower than predicted by the fit to the large distance data points. Recalling that our data points essentially represent a lower limit, it is conceivable that the true pressure is nearer to the power-law predicted curve. In fact, Liu *et al.*[15] showed that weak shock theory was valid quite close to their higher energy sparks, where the pressure excursions were in a similar range to those estimated here. A more fulsome treatment of this experiment would require solution of nonlinear propagation equations [14] and consideration of the full set of plate modes[1,8], but is left for future work.

The setup described in Figs. 2(b-d) suggests a straightforward approach to ACUS characterization of plates. Notably, most previous setups used much larger sensors[3-8] and thus required critical alignment of the source, plate, and receiver in order to isolate the normal-incidence longitudinal waves and to avoid wavelength-scale averaging effects over the sensor surface[5]. Here, the small size of the optomechanical sensor (< 100 μm) implies that normal-incidence waves can be isolated simply by roughly aligning the sensor opposite the spark source. This was confirmed by studies of a variety of glass and plastic plates. For the results described below, the spark-plate distance was set in the 1-2 mm range in order to maximize the SNR. The spark gap was fixed at ~ 2.5 mm, and the arc path was confirmed to lie between the electrodes and parallel to the plate surface (see the SI file). As an aside, we inspected the plates carefully under a microscope and observed no visible signs of damage, even after repeated spark discharges at a single location. All of the data shown was recorded for single-shot (*i.e.,* single spark) pressure events, with no averaging.

Time traces for transmission through two different PDMS plates are shown in Fig. 4. As above, the signal begins with a pressure shock and decaying oscillations at the ~ 2.5 MHz frequency resonance of the sensor. However, the envelope of the signal is modulated by distinct phase and amplitude discontinuities. These discontinuities are periodically spaced, and the period is in good agreement with the round-trip delay times for a pulse circulating inside the PDMS. The dashed vertical lines indicate multiples of $t_{RT} = 2d/v$, where $d$ is the plate thickness and $v = 1030$ m/s is the longitudinal sound velocity for PDMS. Thus, the waveforms, rather complex for the thinner plate, can be interpreted as resulting from interference of multiple sub-reflected blast-wave pulses. The frequency-domain representations of these signals exhibit regularly spaced Fabry-Perot fringes up to several MHz and are provided in the SI file.

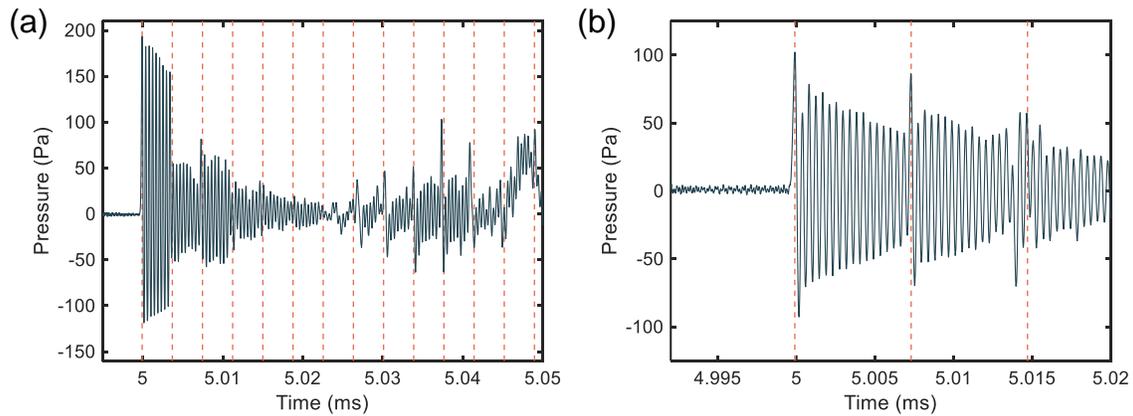

FIG. 3. (a) A typical waveform received on transmission through a 1.9 mm thick PDMS plate. The orange dashed lines indicate multiples of the expected round-trip transit time for the plate (b) As in part a., except for a 3.8 mm thick PDMS plate.

We also studied plates with higher acoustic impedance. In these cases, it was typically more informative to view the data in the frequency domain[4,7,8]. Typical results are shown in Fig. 3, for a 1 mm thick polystyrene petri dish, a 3 mm thick acrylic plate, and a 1 mm thick borosilicate glass slide (see SI file for details). For each case, the FFT algorithm was applied to the received pulse, and the spectrum was normalized by dividing out the intrinsic background spectrum of our sensor[23–25]. Here, the orange dashed lines indicate the predicted longitudinal mode resonance frequencies ($f_m = m/t_{RT}$; $m = 1,2…$). Additional details about the samples, including the sound velocities used, are provided in the SI file. Distinct resonant peaks are observed in each case, riding on a broad background of transmitted energy extending as high as ~ 5 MHz for the plastic samples. Except for one or two extra peaks in Fig. 4(b), excellent alignment between the observed and predicted resonance frequencies is apparent in all cases. It is worth reiterating that these results were obtained without averaging or critical alignment but yielded an SNR similar or better than with conventional setups requiring both[4,5,10]. We hope to report more detailed and quantitative results in future work.

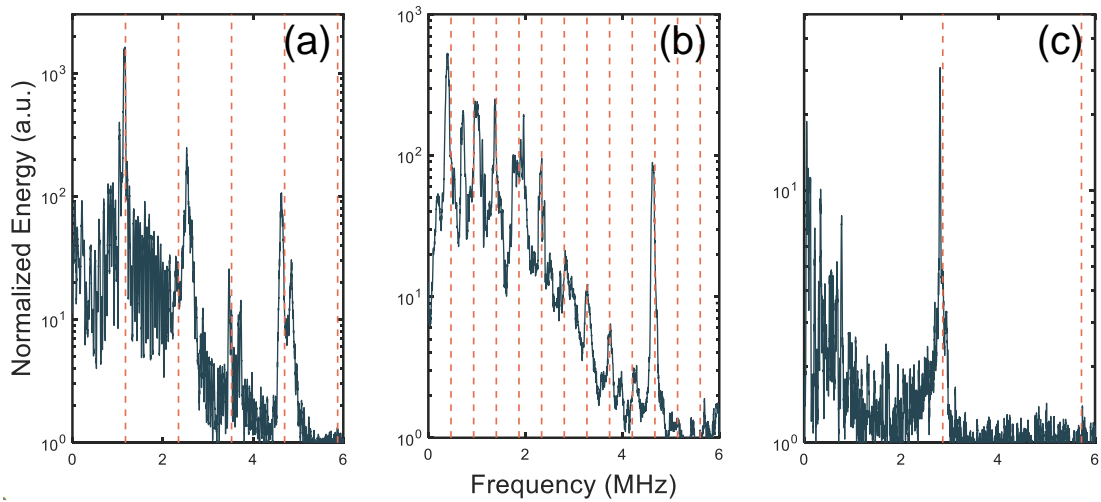

FIG. 4. Normalized energy spectra for (a) ~ 1 mm thick polystyrene, (b) ~ 3 mm thick acrylic, and (c) ~ 1 mm thick borosilicate glass plates. The orange dashed lines indicate the predicted longitudinal Fabry-Perot resonance frequencies in each case. See main text and SI file for additional details.

In summary, we have described an experimental study of the ultrasonic energy produced by low-cost, button-operated piezoelectric spark igniters. These units provide broadband and energetic ultrasound, with bandwidth extending up to at least 5 MHz and peak pressures on the order of 100 kPa near the spark.

In addition, we showed that an optomechanical microphone can be a unique enabler for ACUS setups, due to its high sensitivity, wide bandwidth, and small size. Coupled with the spark source, these properties facilitated a straightforward characterization of a variety of polymer and glass plates, owing to the ease with which longitudinal plate resonances are isolated without critical alignment. This combination of a 'point' source and receiver might present interesting new options for NDT of solid materials.

## SUPPLEMENTARY MATERIAL

See supplementary material for additional information concerning the experimental setup, the optomechanical sensor, the commercial piezoelectric spark igniters, signal processing techniques, physical constants for the plate samples, and additional results.

## ACKNOWLEDGEMENTS

The authors would like to acknowledge the following funding sources which supported this work: Government of Alberta (Innovation Catalyst Grant), Mitacs (Accelerate IT38340), Natural Sciences and Engineering Research Council of Canada (CREATE 495446-17), Alberta EDT Major Innovation Fund (Quantum Technologies).

## REFERENCES

[1] E. Blomme, D. Bulcaen, and F. Declercq, "Air-coupled ultrasonic NDE: experiments in the frequency range 750kHz–2MHz," NDT & E International **35**(7), 417–426 (2002).
[2] D.E. Chimenti, "Review of air-coupled ultrasonic materials characterization," Ultrasonics **54**(7), 1804–1816 (2014).
[3] W. Essig, Y. Bernhardt, D. Döring, I. Solodov, T. Gautzsch, M. Gaal, D. Hufschläger, R. Sommerhuber, M. Brauns, T. Marhenke, J. Hasener, A. Szewieczek, and W. Hillger, "Air-coupled Ultrasound – Emerging NDT Method," Fachbeiträge, 32–43 (2021).
[4] D.A. Hutchins, W.M.D. Wright, and D.W. Schindel, "Ultrasonic measurements in polymeric materials using air-coupled capacitance transducers," The Journal of the Acoustical Society of America **96**(3), 1634–1642 (1994).
[5] D.W. Schindel, and D.A. Hutchins, "Through-thickness characterization of solids by wideband air-coupled ultrasound," Ultrasonics **33**(1), 11–17 (1995).
[6] I. Ladabaum, X. Jin, H.T. Soh, A. Atalar, and B. t. Khuri-Yakub, "Surface micromachined capacitive ultrasonic transducers," IEEE Transactions on Ultrasonics, Ferroelectrics, and Frequency Control **45**(3), 678–690 (1998).
[7] W.M.D. Wright, and D.A. Hutchins, "Air-coupled ultrasonic testing of metals using broadband pulses in through-transmission," Ultrasonics **37**(1), 19–22 (1999).
[8] G. Waag, L. Hoff, and P. Norli, "Air-coupled ultrasonic through-transmission thickness measurements of steel plates," Ultrasonics **56**, 332–339 (2015).
[9] J. Rus, and C.U. Grosse, "Thickness measurement via local ultrasonic resonance spectroscopy," Ultrasonics **109**, 106261 (2021).

[10] K. Bente, J. Rus, H. Mooshofer, M. Gaal, and C.U. Grosse, "Broadband Air-Coupled Ultrasound Emitter and Receiver Enable Simultaneous Measurement of Thickness and Speed of Sound in Solids," Sensors **23**(3), 1379 (2023).

[11] S. Takahashi, "Properties and characteristics of P(VDF/TrFE) transducers manufactured by a solution casting method for use in the MHz-range ultrasound in air," Ultrasonics **52**(3), 422–426 (2012).

[12] R.T. Harrold, in *1980 IEEE International Conference on Electrical Insulation* (1980), pp. 184–189.

[13] W.M. Wright, "Propagation in air of N waves produced by sparks," The Journal of the Acoustical Society of America **73**(6), 1948–1955 (1983).

[14] P. Yuldashev, S. Ollivier, M. Averiyanov, O. Sapozhnikov, V. Khokhlova, and P. Blanc-Benon, "Nonlinear propagation of spark-generated N-waves in air: Modeling and measurements using acoustical and optical methods," The Journal of the Acoustical Society of America **128**(6), 3321–3333 (2010).

[15] Q. Liu, and Y. Zhang, "Shock wave generated by high-energy electric spark discharge," Journal of Applied Physics **116**(15), 153302 (2014).

[16] M.M. Karzova, P.V. Yuldashev, V.A. Khokhlova, S. Ollivier, E. Salze, and P. Blanc-Benon, "Characterization of spark-generated N-waves in air using an optical schlieren method," J Acoust Soc Am **137**(6), 3244–3252 (2015).

[17] P. Yuldashev, M. Karzova, V. Khokhlova, S. Ollivier, and P. Blanc-Benon, "Mach-Zehnder interferometry method for acoustic shock wave measurements in air and broadband calibration of microphones," The Journal of the Acoustical Society of America **137**(6), 3314–3324 (2015).

[18] X. Dai, J. Zhu, and M.R. Haberman, "A focused electric spark source for non-contact stress wave excitation in solids," J Acoust Soc Am **134**(6), EL513 (2013).

[19] J.A. Cooper, R.J. Dewhurst, S. Moody, and S.B. Palmer, "High-voltage spark discharge source as an ultrasonic generator," IEE Proceedings A (Physical Science, Measurement and Instrumentation, Management and Education, Reviews) **131**(4), 275–281 (1984).

[20] C. Ayrault, P. Béquin, and S. Baudin, in *Acoustics 2012*, edited by S.F. d'Acoustique (Nantes, France, 2012).

[21] O.K. Jaenicke, F.G. Hita Martínez, J. Yang, S. Im, and D.B. Go, "Hand-generated piezoelectric mechanical-to-electrical energy conversion plasma," Applied Physics Letters **117**(9), 093901 (2020).

[22] G.J. Hornig, K.G. Scheuer, E.B. Dew, R. Zemp, and R.G. DeCorby, "Ultrasound sensing at thermomechanical limits with optomechanical buckled-dome microcavities," Opt. Express **30**(18), 33083 (2022).

[23] K.G. Scheuer, and R.G. DeCorby, "All-Optical, Air-Coupled Ultrasonic Detection of Low-Pressure Gas Leaks and Observation of Jet Tones in the MHz Range," Sensors **23**(12), 5665 (2023).

[24] K.G. Scheuer, F.B. Romero, G.J. Hornig, and R.G. DeCorby, "Ultrasonic spectroscopy of sessile droplets coupled to optomechanical sensors," (2023).

[25] K.G. Scheuer, F.B. Romero, and R.G. DeCorby, "Observation of the thermal acoustic breathing modes of a bubble," (2023).

[26] S. Dixon, C. Edwards, S.B. Palmer, and J. Reed, "Ultrasonic generation using a plasma igniter," J. Phys. D: Appl. Phys. **34**(7), 1075 (2001).

# *Supplementary Material*

# Air-coupled ultrasound using broadband shock waves from piezolectric spark igniters

**K. G. Scheuer[a] and R. G. DeCorby[b*]**

[a.] Ultracoustics Technologies Ltd., Sherwood Park, AB, Canada T8A 3H5

[b.] ECE Department, University of Alberta, 9211-116 St. NW, Edmonton, AB, Canada, T6G 1H9

*rdecorby@ualberta.ca

1. **Experimental setup**

The opto-acoustic measurement system is similar to that described in our previous work[1,2], and is depicted in Fig. S.1. The optomechanical sensor is interrogated by a laser (Santec TSL-710) tuned to the slope of an optical resonance, and its vibrational motion is imprinted on the reflected signal delivered to a photodetector receiver (Resolved Instruments DPD80). For all of the measurements described in the present work, the average received power at the photodetector was set to ~110 µW. Signals were digitized at 80 MSamples/sec by the RI photoreceiver, and with an internal 40 MHz anti-aliasing filter applied. Digitized waveforms were saved to a computer and subsequently processed in MATLAB.

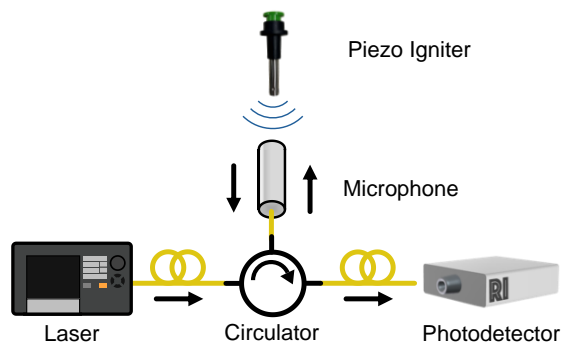

**Supplementary Figure 1.** Schematic representation of the experimental setup. A tunable laser (Santec TSL-710) was connected to an optical circulator and used to interrogate a fiber-coupled optomechanical sensor (Ultracoustics Technologies Broadsonic optical microphone). Reflected light was delivered to a photodetector receiver (Resolved Instruments DPD80). Pressure excursions modulate the reflected laser signal and this is read out as a digitized time-domain electronic signal and subsequently processed in MATLAB.

2. **Optomechanical sensor**

The optomechanical sensor employed here is a 100 µm diameter 'buckled dome' microcavity, very similar to the 'type B' devices we have described in detail previously[1]. However, for the present work the device chip was mounted permanently in a hand-held microphone assembly designed to efficiently couple the sensor to an SMF-28 fiber cable. The sensor itself has a fundamental optical mode linewidth of ~ 1 nm. Given the typical pressure sensitivity of these devices is $\Delta\lambda/\Delta P$ ~ 0.5 nm/kPa[1], it follows that the dynamic range for linear readout (*i.e.*, using a tuned-to-slope technique[1,3]) is on the order of a few hundred Pa, consistent with the results in the main manuscript. It should be noted that this dynamic range limitation is set only by the readout mechanism. In fact, we have subjected these sensors to DC pressures approaching $10^5$ kPa without any apparent damage[4]. Moreover, in the course of this study, we 'accidentally' exposed the microphone to spark blast waves at distances of less than 1 cm on several occasions (*i.e.*, transient pressure excursions $>10^4$ kPa). This resulted in a waveform that was clipped at the hard limits set by the readout technique, but it did not cause any discernible damage to the microphone. A more sophisticated readout technique, such as employing a laser frequency comb[5], might make it possible to operate these sensors with a much higher dynamic range.

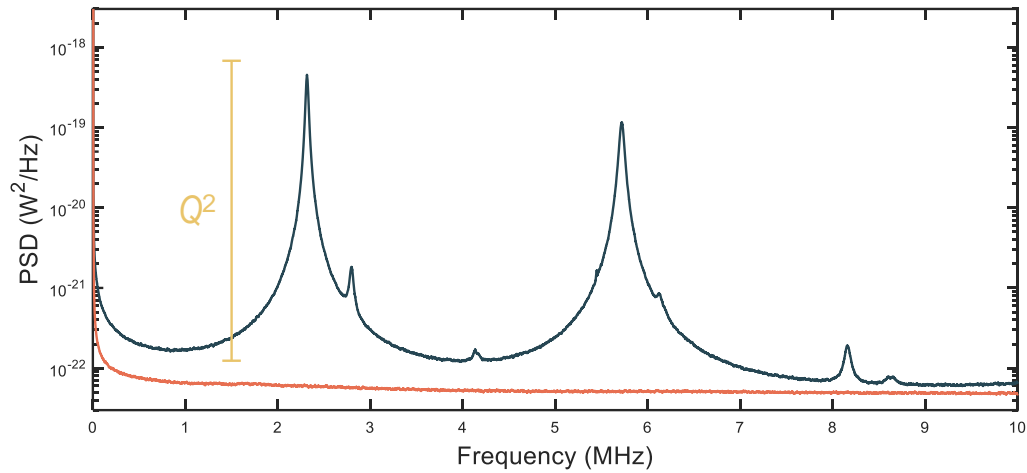

**Supplementary Figure 2.** Thermo-mechanical noise spectrum for the sensor (blue) and shot noise spectrum (orange), at the same time-average receiver power (~ 100 µW) as was used to obtain the results described in the main manuscript.

The thermomechanical noise spectrum for the sensor used is plotted in Fig. S.2, at the same laser power (~110 µW at the photodetector) mentioned above. Also shown is the shot noise level obtained with the laser detuned completely from the optical resonance (*i.e.,* so that that sensor's vibrational energy is not imprinted on the laser) and with the laser power adjusted to result in the same received power at the photodetector. Both of these curves were obtained with averaging of 100 single-shot measurements. As discussed extensively elsewhere[1], and supported by the separation between these noise curves, the optomechanical sensor is very nearly thermo-mechanical noise limited over the entire 0-10 MHz range shown. Further evidence is provided by the fact that the fundamental resonance peak is approximately $Q^2$ times higher than the background noise floor[3], where $Q$ is the mechanical quality factor of the lowest-order mode ($Q \sim 75$ here). This implies that the noise-equivalent pressure (NEP) is almost flat over (most of) this range, since the optomechanical sensor has the same spectral response (*i.e.,* given by its mechanical susceptibility) to both the thermal Langevin force that sets the thermomechanical noise floor and to forces from externally applied signals[1,3,5]. Notably, this range can encapsulate one or more mechanical resonances[6], and extend beyond those resonances. This is quite different than for conventional electrical microphones, for example capacitive membrane-based condenser microphones[7], which can often only be used below their fundamental mechanical resonance. In fact, for the optomechanical microphone operating in a thermomechanical-noise-limited regime, it is possible to normalize a measured spectrum, by dividing it on a point-by-point basis with the background noise spectrum[2,6]. This essentially 'removes' the intrinsically resonant response of the sensor, so that the signal spectrum can be assessed separately.

Importantly, such a deconvolution only works provided the thermomechanical noise is substantially higher than the background (shot) noise level. For the present sensor, as shown in Fig. 2, this is only approximately true. Shot noise likely plays a minor but non-negligible role over the entire frequency range, and laser-intensity noise and electronic 1/f noise cannot be neglected at frequencies below ~ 100 kHz. It is also worth noting that laser noise can be greatly reduced by using differential detection[6].

To roughly calibrate our sensor, we used acoustic tones (generated by a smart phone speaker) in the audio range. Our sensor and a commercial ultrasonic microphone (Dodotronic Ultramic 384K BLE) were placed at the same distance from the tone source, so that the built-in sound-pressure-level (SPL) calibration provided with the commercial microphone could be used to translate our optical power signals to pressure. This low-frequency calibration can be translated to higher frequencies by taking into account the device mechanical susceptibility, as discussed above. Reassuringly, this calibration procedure resulted in a noise-equivalent-pressure estimate for our sensor ($NEP$ ~ 50-70 $\mu Pa/Hz^{1/2}$; see for example Fig. 1(b) in the main manuscript) which is in excellent agreement with a more sophisticated calibration from our earlier work[1].

3. **Piezoelectric spark igniters**

The results from Fig. 1 of the main manuscript (and for the yellow symbols in Fig. 2(a) of the main manuscript) were obtained using a particular brand of camp-stove igniters ('Optimus Sparky'; see Fig. S.3(a)). Other brands were tested, and produced similar blast wave pulses, but the Sparky units were found to have superior spark-to-spark and unit-to-unit consistency in terms of the shape and amplitude of the acoustic signal. For through-plate measurements, we found it more flexible and convenient to use a barbecue replacement part ('Master Chef OEM Piezo Igniter') designed to generate a spark between externally wired electrodes and shown in Fig. S.3(b).

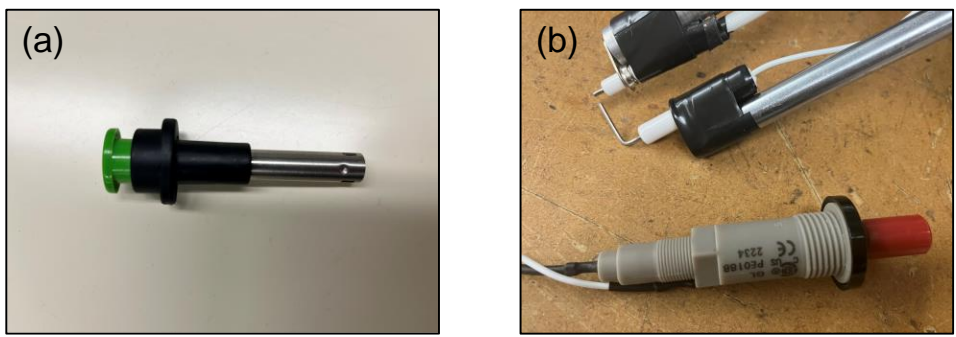

**Supplementary Figure 3.** (a) A photograph of an 'Optimus Sparky' piezoelectric campstove igniter is shown. (b) A photograph of the 'Master Chef' OEM replacement barbeque igniter, including the electrodes wired out from the terminals of the button-operated piezoelectric spark unit.

For the plate measurements, the electrodes provided with the Master Chef igniter were simply mounted to generate a spark across a gap set to a similar value (~ 2-3mm) as the pin-to-barrel gap distance in the campstove igniters. We verified first that this resulted in a very similar blast waveform as for the campstove igniter, at distances of ~ 10 cm or greater in air, including few-microsecond duration N-wave pulses. These electrodes were then positioned within a few mm of a plate surface, with the microphone positioned directly opposite and across the plate and typically at an angle to minimize the possibility of acoustic reverberations between the plate and the microphone. As shown in Fig. S.4, when sparks were generated near glass and polymer plates, the arcing was observed to occur predominately between the metal electrodes and parallel to the plate surface. There was spark-to-spark variation in the termination points of the electrical arc, which we estimated to be on the order of +/- 0.5 mm in the direction normal to the plate surface. At very close spark-plate distances on the order of 1 mm, we could not rule out the possibility of some arcing occurring directly to the plate.

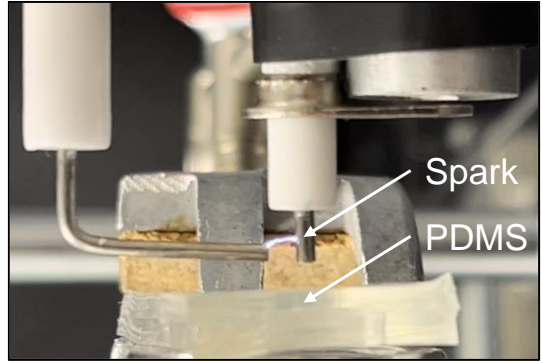

**Supplementary Figure 4.** Photograph of a spark generated adjacent to a PDMS plate shown in cross-section, ~3.8 mm thick. The spark was generated using the externally wired piezo igniter from Fig. S.3(b). The spark gap is ~ 2.5 mm here, and the path of the spark is approximately 3.5 mm from the front plate surface. With this electrode arrangement, the spark end points were observed to vary between sparks by as much as ~ 1 mm in the direction normal to the plate. For the results plotted in Fig. 2(a) of the main manuscript, the spark-plate distance was measured to the center of the horizontal (left) electrode in the figure above, but variations in the spark path mentioned result in some uncertainty in the actual spark-to-plate distance for a given spark event.

## 4. Signal processing

As mentioned above, our measurement scheme utilizes a 'tuned-to-slope' technique where the interrogation laser is slightly detuned from the fundamental optical resonance of the sensor. Changes in cavity length caused by thermal vibrational energy or by incoming pressure waves are thus imprinted on the reflected optical power, and these fluctuating time-domain signals are converted to an electrical signal by the photodetector receiver.

For the present work, the pressure excursion caused by a spark blast wave was used to trigger the capture of a time record (typically 10 msec in total duration). Portions of representative time traces are shown in Fig. 1(a), Fig. 2(d), and Fig. 3 of the main manuscript. All of these are 'single shot' (*i.e.*, single spark event) data captures without averaging or smoothing applied, indicative of the excellent SNR provided by the optomechanical sensor. A fast-Fourier-transform (FFT) algorithm (in MATLAB) was applied to these time traces to produce the frequency-domain energy spectra shown in the main manuscript. In all cases, the time trace was truncated after the main pulse arrival, to remove the impact of spurious reverberations caused by the microphone housing and acoustic reflections within the sensor substrate[5]. In Fig. 1(b) of the main manuscript, the FFT was calculated using either the raw data up to and including the initial N-pulse or using only the zero-padded N-Pulse. In Fig. 4 of the main manuscript, raw spectra were generated in a similar way (*i.e.*, using the truncated time-domain signal without zero-padding), for both a section of the time trace containing the pulse transmitted through the plate and for a section for the time trace prior to the pulse arrival (*i.e.*, background noise only). These spectra were gently smoothed using the 'movmedian' function with a 100-point window in MATLAB, and then the signal spectrum was divided by the noise spectrum on a point-by-point basis to produce the normalized energy density plots shown.

## 5. Plate samples studied

Table S.1. lists the properties of the plate samples used to obtain the results in Figs. 3 and 4 of the main manuscript. The thickness in each case was estimated using a micrometer, and the speed of sound and acoustic impedance were based on values reported in the literature as cited in the table.

| Plate material | Acoustic impedance (Rayl) | Speed of sound (m/s) | Plate thickness (mm) |
|---|---|---|---|
| PDMS (fabricated in house)[8,9] | $1.05 \times 10^6$ | 1030 | 1.9 and 3.8 |
| Polystyrene (petri dish)[10] | $2.47 \times 10^6$ | 2350 | 1.0 |
| Acrylic (black PMMA)[10] | $3.25 \times 10^6$ | 2730 | 3.0 |
| Soda-lime glass (Fisher Scientific microscope slide)[11] | $14.2 \times 10^6$ | 5720 | 1.0 |

## 6. Additional results

### 6.1. Positive pressure excursion captured at closer distance in air

As mentioned in the main text, it was possible to bias the microphone such that a larger positive pressure excursion could be captured from a nearby spark event. Essentially, this involves setting the laser wavelength to be aligned near a wing of the fundamental optical resonance, rather than near the midway point of the Lorentzian lineshape[1,3]. In this case, the positive compression phase of the blast wave is captured without clipping while the negative rarefaction phase is strongly clipped at the limit set by the off-resonance reflectance of the optomechanical sensor. A typical result is shown in Fig. S.5, where the sensor was placed ~ 6 cm from the spark produced by the 'Optimus Sparky' piezoelectric campstove igniter.

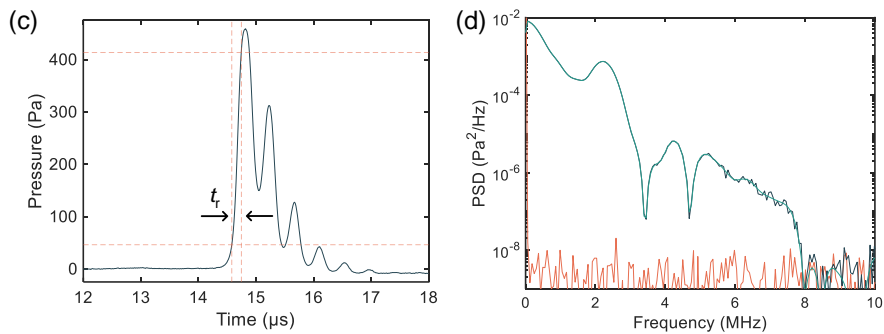

**Supplementary Figure 5.** (a) A typical positive pressure excursion measured at ~ 6 cm spacing, with the 10-90% rise time ($t_r$ ~ 0.16 μs) indicated. The oscillations with period ~0.4 μs arise from the sensor fundamental resonance at ~2.5 MHz. (b) Frequency spectrum for the pulse in part a (blue/green) along with the noise floor (orange), both calculated from shorter time samples (*i.e.* lower FFT

resolution) compared to the results in Fig. 1 of the main manuscript. Frequency content extending up to nearly 8 MHz is estimated for the positive pressure 'pulse'.

It should be noted first that this trace undoubtedly contains some nonlinear distortion, since it is captured using a nearly complete valley-to-peak transition of the Lorentzian lineshape, which is not well approximated as a straight-line slope near its extremities. Nevertheless, we were careful to avoid clipping of the positive excursion. From this type of data, a few insights can be gleaned:

i. The rise time of the front shock was consistently observed to be in the 100-200 ns range, in good agreement with expectations based on the literature (see main text).
ii. At this distance, the spark delivers sufficient energy in the 2.5 MHz range to excite a distinct 'ringing' of the sensor's fundamental resonance.
iii. The positive pulse has above-noise frequency content extending up to at least 5 MHz, consistent with the plate measurement data from the main manuscript.

6.2. Supplementary plate measurement data

Fig. S.6 shows The FFT plots calculated from the time-domain traces shown in Fig. 3 of the main manuscript. Consistent with the clear signatures of periodic pulse arrivals in the time-domain plots, the frequency spectrum contains regularly spaced Fabry-Perot fringes with remarkably high SNR in the ~ 0 – 2 MHz range, and visible up to and beyond 4 MHz for the thinner sample.

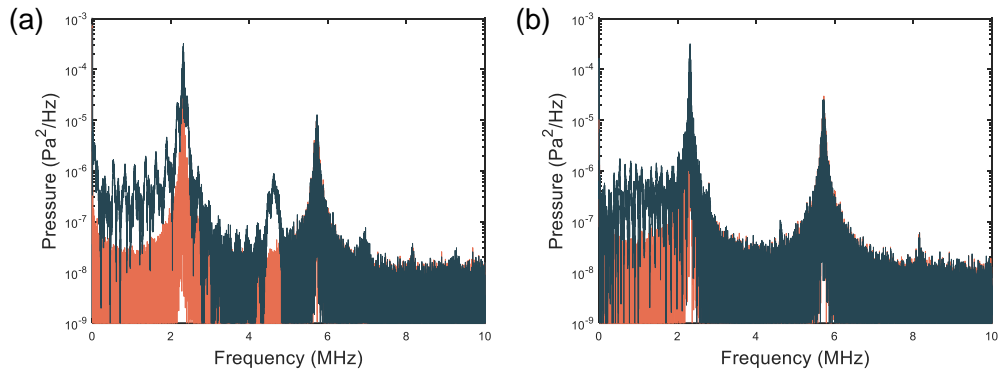

**Supplementary Figure 6.** Frequency-domain FFT plots for the time-domain traces shown in Fig. 3 of the main manuscript, for (a) ~ 1.9 mm thick PDMS and (b) ~3.8 mm thick PDMS. In each plot, the blue curve is the FFT including the ultrasound pulse while the orange curve is the FFT excluding that feature (*i.e.*, background thermomechanical noise).

The time-domain traces used to generate the FFT spectra in Fig. 4 of the main manuscript are shown in Fig. S.7. These are very typical examples of the transmitted portion of the spark's blast-wave energy for each of these plate materials. However, we found that waveforms were somewhat sensitive to the spacing and angle between the microphone and the plate, likely due to reverberations between in the air gap between them. To minimize this effect, the waveforms shown were captured with the sensor at ~ 5 mm from the plate surface and with the microphone angled relative to the plate. They are all characterized by a rapid onset of pressure (rise time ~ 100 ns), followed by a distinct ringing at the sensor's fundamental resonance frequency. Typically, a few hundred microseconds after the initial pulse arrival, the waveform would become dominated by lower-frequency content which we attributed to energy arriving at the microphone from other paths (*i.e.,* 'leaking' around the plate). For FFT calculations, such as which produced the data in Fig. 4 of the main manuscript (and see below), the signals were truncated to exclude these portions of the time-domain waveform.

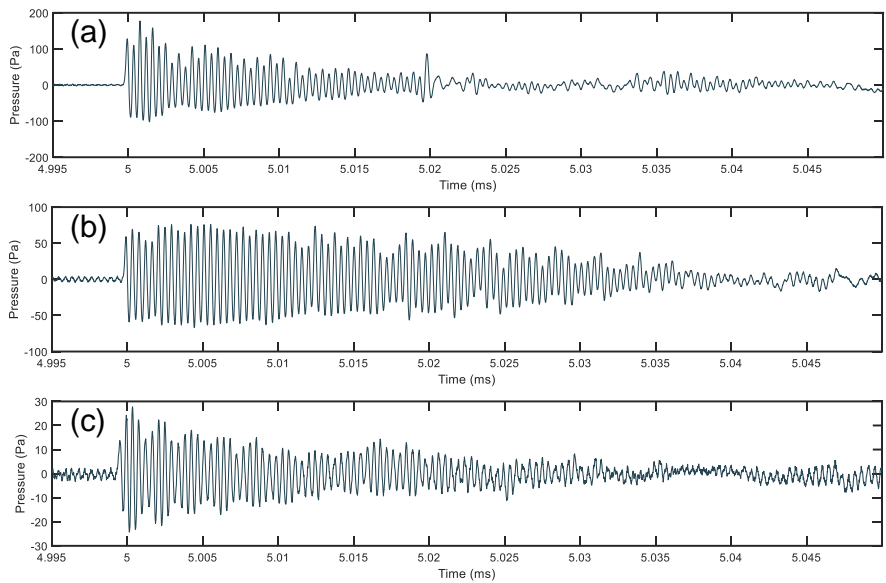

**Supplementary Figure 7.** Time-domain waveforms for the frequency spectra shown in Fig. 4 of the main manuscript, corresponding to (a) ~ 1 mm polystyrene, (b) ~ 3 mm acrylic, and (c) ~ 1 mm soda lime glass.

Representative 'raw' FFT spectra, for the pulses in Fig. S.7(a) and (b), are shown in Fig. S.8. Here, the orange traces are single-shot (non-averaged) background noise spectra calculated using a portion of the time waveform prior to pulse arrival, and the green traces are the spectra calculated using a trace of the same duration but including the portion of the pulse prior to arrival of 'leaked' signal as described above. These raw data sets support the conclusion from the main manuscript, and above, that significant energy content up to at least 5 MHz is present in the transmitted pulses. To produce the graphs shown in Fig. 4 of the main manuscript, non-averaged traces of this type were slightly smoothed using a the 'movmedian' function in MATLAB (100-point window), and the signal spectrum was divided by the noise spectrum, on a point-by-point basis.

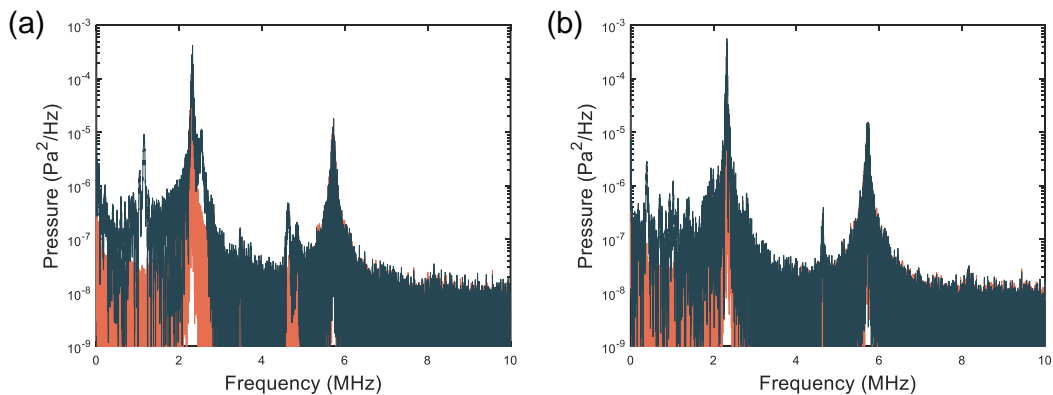

**Supplementary Figure 8.** Raw spectra for the time-domain signals from Figs. S.7(a) and (b). The normalized versions of these plots (*i.e.*, the blue curve divided by the orange curve) are shown in Figs. 4(a) and (b) of the main manuscript.

**Supplementary References:**

[1] G.J. Hornig, K.G. Scheuer, E.B. Dew, R. Zemp, and R.G. DeCorby, "Ultrasound sensing at thermomechanical limits with optomechanical buckled-dome microcavities," Opt. Express **30**(18), 33083 (2022).

[2] K.G. Scheuer, and R.G. DeCorby, "All-Optical, Air-Coupled Ultrasonic Detection of Low-Pressure Gas Leaks and Observation of Jet Tones in the MHz Range," Sensors **23**(12), 5665 (2023).

[3] B.-B. Li, L. Ou, Y. Lei, and Y.-C. Liu, "Cavity optomechanical sensing," Nanophotonics **10**(11), 2799–2832 (2021).

[4] S. Al-Sumaidae, L. Bu, G.J. Hornig, M.H. Bitarafan, and R.G. DeCorby, "Pressure sensing with high-finesse monolithic buckled-dome microcavities," Appl. Opt., AO **60**(29), 9219–9224 (2021).

[5] W.J. Westerveld, M. Mahmud-Ul-Hasan, R. Shnaiderman, V. Ntziachristos, X. Rottenberg, S. Severi, and V. Rochus, "Sensitive, small, broadband and scalable optomechanical ultrasound sensor in silicon photonics," Nat. Photonics **15**(5), 341–345 (2021).

[6] F. Zhou, Y. Bao, R. Madugani, D.A. Long, J.J. Gorman, and T.W. LeBrun, "Broadband thermomechanically limited sensing with an optomechanical accelerometer," Optica, OPTICA **8**(3), 350–356 (2021).

[7] T.B. Gabrielson, "Mechanical-thermal noise in micromachined acoustic and vibration sensors," IEEE Transactions on Electron Devices **40**(5), 903–909 (1993).

[8] R.-M. Guillermic, M. Lanoy, A. Strybulevych, and J.H. Page, "A PDMS-based broadband acoustic impedance matched material for underwater applications," Ultrasonics **94**, 152–157 (2019).

[9] G. Xu, Z. Ni, X. Chen, J. Tu, X. Guo, H. Bruus, and D. Zhang, "Acoustic Characterization of Polydimethylsiloxane for Microscale Acoustofluidics," Phys. Rev. Applied **13**(5), 054069 (2020).

[10] G. Destgeer, J.H. Jung, J. Park, H. Ahmed, K. Park, R. Ahmad, and H.J. Sung, "Acoustic impedance-based manipulation of elastic microspheres using travelling surface acoustic waves," RSC Adv. **7**(36), 22524–22530 (2017).

[11] K.T. Son, and C.C. Lee, "Bonding and impedance matching of acoustic transducers using silver epoxy," Ultrasonics **52**(4), 555–563 (2012).